\documentclass[aps,prb,preprint,superscriptaddress,showpacs,floatfix]{revtex4}

\usepackage{graphicx}
\begin{document}

\title{Metastable Resistivity States and Conductivity Fluctuations in Low-doped La$_{1-x}$Ca$_{x}$MnO$_3$ Manganite Single Crystals}

\author{B. Dolgin}
\affiliation{Department of Physics, Ben Gurion University of the Negev, P.O. BOX 653,
84105 Beer Sheva, Israel}

\author{M. Belogolovskii}
\affiliation{Donetsk Institute for Physics and Engineering, National Academy of Sciences of Ukraine, 83114 Donetsk, Ukraine}

\author{X. D. Wu}
\affiliation{Department of Physics, Ben Gurion University of the Negev, P.O. BOX 653,
84105 Beer Sheva, Israel}
\affiliation{Department of Materials Engineering, Monash University, Clayton, Australia, 3800}

\author{V. Markovich}
\affiliation{Department of Physics, Ben Gurion University of the Negev, P.O. BOX 653,
84105 Beer Sheva, Israel}

\author{G. Jung}
\affiliation{Department of Physics, Ben Gurion University of the Negev, P.O. BOX 653,
84105 Beer Sheva, Israel}

\date{\today}

\begin{abstract}

Conductivity noise in dc current biased La$_{0.82}$Ca$_{0.18}$MnO$_{3}$ single crystals has been
investigated in different metastable resistivity states enforced by applying voltage pulses to the sample at low temperatures. Noise measured in all investigated resistivity states is of 1/f-type and its intensity  at high temperatures and low dc bias scales as a square of the bias. At liquid nitrogen temperatures for under bias exceeding
a threshold value, the behavior of the noise deviates from above quasi-equilibrium modulation
noise and  depends in a non monotonic way on applied bias. The bias range of nonequilibrium 1/f noise coincides with the range at which the conductance increases linearly with bias voltage. This feature is attributed to a broad continuity of states enabling indirect inelastic tunneling across intrinsic tunnel junctions. The nonequilibrium noise has been ascribed to indirect intrinsic tunneling mechanism while resistivity changes in metastable states to variations in the energy landscape for charge carriers introduced by microcracks created by the pulse procedures employed.
\end{abstract}

\maketitle

\section{Introduction}

	Metastability is a generic feature of phase separated systems with two or more competing ordering mechanisms. Mixed valence manganites are characterized by a complex phase diagram containing many magnetic and electronic phases. The multi-phase state stems from an interplay of structural, charge, orbital, and spin degrees of freedom with comparable energy scales. Phase separation (PS), is claimed to be responsible for most of the peculiar properties of perovskite manganites, including the very appearance of the colossal magnetoresistive effect (CMR).\cite{dagotto} Double exchange (DE) interactions are considered to be responsible for ferromagnetic ordering in La$_{1-x}$Ca$_x$MnO$_3$ (LCMO) manganites below $T_C$. However, in LCMO doped below the percolation threshold, $x<x_c=0.225$ \cite{okuda}, a ferromagnetic insulating phase incompatible with the DE mechanism appears.\cite{papa,papa2} Several experimental  and theoretical papers concluded that transport and magnetic properties of low-doped manganites are governed by superexchange and orbital ordering acting hand in hand with DE interactions.\cite{okuda,papa,vanacken,tokura} Presence of two ferromagnetic phases with different orbital order and electronic properties\cite{papa,papa2,henion} leads to appearance of metastable states with markedly different resistivity.\cite{APL018,PRB018,PRB020,EJPB020,PRB020films,gao,levy,mathieu,anane} Metastable states are characterized by history dependent conductivity, pronounced relaxation of magnetization and resistivity, memory effects in electric resistance and magnetization, and strong 1/f-type conductivity noise frequently accompanied by non-Gaussian  fluctuations.

	Experimentally, metastable states can be induced by making relatively fast changes of experimental parameters such as magnetic field, electrical bias, or temperature. The majority of reported experiments on metastable states in manganites has been based on investigations of relaxation processes appearing after abrupt changes of the applied magnetic field.\cite{levy,mathieu,anane} We have for some time concentrated on investigations of metastable states induced by electric field/current procedures.\cite{APL018,PRB018,PRB020,EJPB020,PRB020films} We have used both continuous current sweeps and short current pulses to create low and high resistivity metastable states in LCMO single crystals and thin films. While relatively slow current sweeps with progressively increasing amplitude tend to decrease sample resistivity, application of bias pulses at low temperatures have much stronger effect and increases the resistivity  by orders of magnitude. Current enforced low and high metastable resistivity  states survive thermal cycling to room temperatures and are characterized by long lifetimes, significantly exceeding the time scale of an experiment. In the experiments performed with freshly fabricated crystals, current induced states could have been relatively easily rejuvenated to the initial pristine state by means of thermal treatment at slightly elevated temperatures (around 400 K). However, with elapsing time, LCMO samples spontaneously evolve towards higher resistivity states and it becomes progressively more difficult to enforce metastable states and/or to rejuvenate them. \cite{EJPB020} Such ageing effects impose still not fully answered question about the ultimate equilibrium state of low-doped manganites.

Strong and weak external stimuli create different metastable states in low-doped LCMO. The dominant effect of the weak currents may consist in reversible injection of spin-polarized carriers into specific ferromagnetic domains. Charges accumulated at phase boundaries of ferromagnetic metallic regions can be driven by electric field/current and literally pull the boundaries into insulating regions, thus rising the volume occupied by the metallic phase.\cite{CER1} On the other hand, strong pulses most likely act through the associated strong electric fields and induce local changes to the orbital order in a less conducting phase and/or change the overall topology of the insulating phase configuration.	Metastable resistivity states are closely related to the effects of resistivity controlled by electric currents in so-called colossal electroresistive effect (CER).\cite{APL018,CER1,CER2,CER3,CER4,CER5} The mechanism that accounts for CER, whether attributed to spin polarized currents or to conductance channels, is not yet fully understood. The nature of charge transport in metastable states of colossal magnetoresistance manganites remains therefore still an open question, despite a significant effort devoted to that subject.

Micro scale phase separation in CMR manganites is self-organized by intrinsic energy landscapes containing hierarchical energy barriers for relieving of the strain.\cite{ahn,Bishop} Electronic, magnetic, and structural inhomogeneities, native to mixed valence manganites and other transition metal oxides, are associated with elastic fields of long range strain effects due to cracks and defects. Presence of cracks in single crystalline manganite samples affect their resistive behavior.\cite{V3} Application of a sufficiently strong electric current/field may induce strain resulting in microcracks and shift domain boundaries in a metastable pinning landscape. Domain walls pinned at boundaries between ferromagnetic phases with different orbital order can be directly coupled to the strain fields in the sample. External stimuli may influence the topology of the coexisting metallic and insulating regions and cause significant changes in the transport properties. Thermal rejuvenation to the pristine state occurs at temperatures at which thermal fluctuations exceed the pinning energy of the hierarchical pinning landscape. One may assume therefore that metastable resistivity states in low doped manganites are enforced by changes in the energy landscape of charge carriers.

Fluctuations in condensed-matter systems arise from charge carriers' relaxation processes strongly influenced by the underlying energy landscape. Therefore, changes in the energy landscape should be directly reflected in  noise properties of various metastable resistivity states. In this paper we discuss experiments employing conductivity noise as a tool for investigations of metastable resistivity states associated with microcracking in La$_{0.82}$Ca$_{0.18}$MnO$_3$ manganite single crystals.

\section{Theoretical background}

Low frequency noise is the major factor limiting the performance of practical devices fabricated from doped manganites. Basic noise properties of manganites have been investigated in this context. However, only in-depth studies, beyond a simple estimate of the magnitude of the noise level and signal-to-noise ratio, provide a tool for understanding the dynamics of transport and magnetization processes and may also provide a unique tool for understanding the nature and dynamics of metastable resistivity states.  Since the "rediscovery" of mixed valance manganites in the early 90`s, many publications addressed the issue of noise (for an overview of recent developments see e.g., an introduction paragraph to our recent paper  ~\onlinecite{JAP}). Almost all investigations revealed prominent broad band conductivity fluctuations with a power spectral density (PSD) following a 1/f$^\alpha$ law, with the spectral exponent $\alpha$ equal or close to 1.

Usually, 1/f noise spectra in condensed-matter systems come from assembles of fluctuators with well defined characteristic rates.\cite{DH,weissrev} When elementary fluctuators are thermally activated two-state Markov systems, the noise spectrum from each has a Lorentzian form
\begin{equation}
S_i(\omega)\propto \frac{\tau}{1+(\omega\tau)^2},
\end{equation}
where $\tau=\tau_0\exp(E/k_BT)$ is the characteristic relaxation rate. The resulting total noise spectrum of an ensemble
  \begin{equation}
  S(\omega)=\int{S_i(\omega,\tau)D(\tau)}d\tau\propto \int{\frac{\tau_0\exp[E/k_BT]}{1+\omega^2\tau^2\exp[2E/k_BT]} D(E)dE }
    \end{equation}
has 1/f form for a flat distribution of activation energies $D(E)=const.$  Dutta and Horn, (DH) have shown that 1/f$^\alpha$ spectra with $\alpha\approx 1$ arise not only for $D(E)=const.$ but also for the distribution function $D(E)$ that does not vary much in the energy range $|kT_B\ln(\tau_0/\omega_1)|<E< |k_BT\ln(\tau_0/
\omega_2)|$, where $\omega_1$ and $\omega_2$ are the limits of the observable 1/f noise frequency range. \cite{DH,vanZiel}  When the product $S(\omega)\omega$  is a weak function of frequency, then Dutta Horn equation
\begin{equation}
\alpha=-\frac{\partial \ln S(\omega,T)}{\partial \ln \omega} =1-\frac{1}{\ln\left(\omega\tau_0\right)}\left(\frac{\partial \ln S(\omega,T)}{\partial \ln T}-1 \right).
\end{equation}
within the limits discussed in detail in ref. ~\onlinecite{weissrev}, can be used to infer approximate attempt rates $\tau_0$  and activation energy distributions $D(E)$ :
\begin{equation}
\frac{S(\omega,T)\omega}{k_BT}\propto D(\tilde{E}=k_BT\ln(\omega\tau_0)). \label{PSD-D}
\end{equation}

When the  observed noise obeys DH model then equation (2) enables an insight into the underlying energy distributions $D(E)$. A flat $D(E)$ distribution, $\partial D(E)/\partial E= 0$, gives rise to a pure 1/f spectrum with the spectral exponent $\alpha=1$ and a linear temperature dependence of the noise level. Departure from the linearity with spectral exponent $\alpha\neq1$ implies a non-zero $\partial D(E)/\partial E$. When $\alpha < 1$ or $\alpha > 1$, then there is an excess in the density of the low, or respectively high energy fluctuators and $\partial D(E)/\partial E$ derivative is negative/positive within the experimental energy window.

Conductivity noise with $1/f$ spectrum is generally related to resistance fluctuations which are measured by applying dc current and recorded as voltage fluctuations. When the resistance fluctuations are  not influenced by its flow,  by the current flow, but only converted by flowing current into measurable voltage fluctuations, then PSD of the noise scales as the square of the bias current. Such modulation noise of equilibrium resistivity fluctuations is  referred to in the literature as "equilibrium $1/f$ noise", see e.g. references [\onlinecite{eqnoise1,eqnoise2,eqnoise3}]. Of course this does not mean that the sample is in the state of total thermodynamic equilibrium concerning its microstate. There is a mounting experimental evidence that $1/f$ noise of equilibrium conductivity fluctuations in many systems is accompanied by the noise directly modified by, the passage of current through the sample. Recent examples of such noise in doped manganites was described in references [\onlinecite{APL018,barone,nowak}]. This type of noise is referred to in the literature  as  $1/f$ noise of nonequilibrium conductivity fluctuations, or in brief as to "nonequilibrium 1/f noise". Such noise is characterized by bias dependence which is considerably different from the quadratic power law.\cite{eqnoise2}

Recently, we have intensively investigated conductivity noise of La$_{0.82}$Ca$_{0.18}$MnO$_3$ manganite crystals.\cite{JAP,APL} At high temperatures, La$_{0.82}$Ca$_{0.18}$MnO$_3$ is in a paramagnetic insulating state, and the resistivity, dominated by a hopping mechanism, increases with decreasing temperature, $dR=dT < 0$. The resistivity reaches a pronounced maximum, related to the metal-insulator transition, at a point very close to the paramagnetic-to-ferromagnetic transition at $T_C$. In the ferromagnetic state below $T_C$, the intrinsic phase separation leads to percolation, metallic-like conductivity with $dR/dT > 0$. However, with further temperature decrease, the resistivity reaches a strong upturn at temperatures below 100 K, associated with increased concentration of the ferromagnetic insulating phase and domination of the conductivity by intrinsic tunneling mechanism.\cite{PRB018}

We have demonstrated that the dominant conductivity noise at all investigated temperatures, despite changing dissipation mechanism and magnetic state of the system, is of 1/f-type noise and obeys DH model predictions.\cite{JAP} At all temperatures, with exception of temperatures below the resistivity upturn in the $R(T)$ characteristics the noise has an equilibrium character, in the sense that conductivity fluctuations do not depend on applied bias. However, at low temperatures, where intrinsic tunneling dominates the conductivity, the spectral density of the noise initially increases proportionally to $I^2$, as expected for current independent resistivity fluctuations, but at higher currents the noise becomes dominated by bias dependent nonequilibrium resistivity fluctuations, what manifests itself in deviations of the bias dependence of the spectral density from the power law $S_V\propto I^2$. In our previous work we have eliminated experimental artifacts as a possible reason for this esoteric behavior and tentatively ascribed it to bias induced changes in the tunneling mechanism which are correlated with changes in the underlying energy landscape.\cite{APL,JAP}

Observe that 1/f noise obeying DH model originates from superposition of large number of elementary fluctuators. Such model meets the assumptions of the central limit theorem and thus predicts the Gaussian noise.\cite{Nelkin} Surprisingly, significant non-Gaussian 1/f noise has been recently observed in La$_{0.80}$Ca$_{0.20}$MnO$_3$ manganite single crystals at low temperatures.\cite{hodu} The non-Gaussian character of the fluctuations was demonstrated through measurements of the probability density function and second spectra (fourth moment) measurements. It was observed that the noise becomes non-Gaussian when the material is cooled down below the ferromagnetic transition temperature $T_C$. With further cooling deeper into the phase separated state, the noise becomes even more non-Gaussian. Moreover, the observed temperature dependent 1/f noise magnitude shows a sharp freeze out with temperature on cooling into very low temperatures. The authors proposed that the non-Gaussian noise arises from charge fluctuations in a correlated glassy phase of the polaronic carriers which develop in these systems according to numerical simulation studies. Let us underline that the regime of appearance of non-Gaussian 1/f noise is the same one at which we have reported appearance of the "nonequilibrium 1/f noise". However, the major difference is that non-Gaussian character of the noise was revealed in almost equilibrium conditions using a very small excitation with ac bias current while the "nonequilibrium noise" appears at strong dc bias. Since we did not performed yet measurements of the higher moments, it is still unclear to us whether the two phenomena appear simultaneously and to which extend their physical mechanisms are related.

At lowest temperatures studied in the experiment, charges in the investigated system are transmitted by tunneling across intrinsic barriers which arise due to phase separation. When the decay length of the electron wave functions exceeds the thickness of the intrinsic tunnel barrier then direct  the conduction is dominated by elastic tunneling mechanism. With increasing thickness of the tunneling  barrier, hopping along chains of localized states becomes more favored than direct inelasting tunneling. While hopping along localized states path, a carrier does not cross quantum-mechanically the entire distance between the electrodes, but rather jumps from the junction electrode to the first state, lose the phase memory,  moves to the second nano-island and, eventually, after completing all the hopping path, jumps to the opposite electrode. Such case was considered by Glazman and Matveev~\cite{GM,beasley} in the model (GM) of indirect tunneling in disordered materials.  The GM model was found to apply very well to low temperature transport in perovskite manganites.\cite{PRB018,gross,bertina}

GM model predicts that  at low temperatures the voltage dependencies of the tunnel conductance reflects multistep tunneling via $N$ localized states:
\begin{equation}
G(V)=G_0+\sum_{N=1}^{\infty}G_N(V,T),
\end{equation}
where $G_0$ represents bias and temperature independent elastic tunneling term, while $G_N$ describe tunneling through $N\geq1$ localized states.
\begin{eqnarray}
G_N(V)=a_NV^{(N-\frac{2}{N+1})} \hskip 1truecm\mbox{for}\hskip 0.3truecm eV\gg k_BT,\\
G_N(T)=b_NT^{(N-\frac{2}{N+1})} \hskip 1truecm\mbox{for}\hskip 0.3truecm eV\ll k_BT,
\end{eqnarray}
where coefficients $a_N$ and $b_N$ depend exponentially on barrier thickness.

The differential conductance of a thin, inhomogeneous, insulating barrier is therefore a power function of the voltage, see also Ref.~\onlinecite{CEJP},
\begin{equation}
G_{\rm{in}}(V) = G_0+{\rm{const}}\cdot V^n \label{power},
\end{equation}
where index $n$ characterizes the tunneling regime: $n=2$ corresponds to elastic tunneling with the energy relaxation in the conducting regions of the system, whereas other $n$ are signatures of inelastic tunneling in which an electron losses its energy inside the insulating region. By finding the value of index $n$ from experimental data one can infer information about the physics of electron transport processes across the dielectric layer.

Results of our previous experiments have leaded us to the hypothesis that the appearance of "nonequilibrium 1/f" noise is associated with changes in the mechanism of intrinsic tunneling which dominates transport properties of lightly doped LCMO manganites at low temperatures.  We have noted that that metastable states are most efficiently enforced by field/current procedures exactly at the same temperatures at which the "nonequillibrium 1/f noise" appears. However, our initial experiments did not addressed the issue of "nonequilibrium 1/f noise" in metastable resistivity states.  If the above conclusions are correct, it may be predicted that properties of the noise, and in particular of "nonequillibrium 1/f noise", should be markedly different in different metastable states.

\section{Experimental and results}

La$_{0.82}$Ca$_{0.18}$MnO$_3$ crystals were grown by a floating zone method using radiative heating.\cite{crystal} X-ray data of the crystal were compatible with the perovskite structure orthorhombic unit cell, $a = 5.5062$ \AA,  $b = 7.7774$ \AA, $c = 5.514$ \AA. The as-grown crystal, in form of a cylinder, about 4 cm long and 4 mm in diameter, was cut into individual small rectangular 6 $\times$ 3 $\times $ 2 mm$^3$ bars, with the longest dimension along the $<110>$ crystallographic direction. Current and voltage leads were indium soldered to gold/chromium contacts deposited by thermal evaporation in vacuum.

For noise measurements the sample was thermally anchored to the sample holder of a variable temperature liquid nitrogen cryostat. Conductivity noise was measured in a conventional 4-point contact arrangement  by biasing the sample with dc current, supplied by high output impedance current source, and measuring the resulting voltage fluctuations. The voltage signal was amplified by a home made room temperature low noise preamplifier, located at the top of the cryostat, and further processed by a computer assisted digital signal analyzer. To eliminate environmental interferences and noise contributed by the measuring chain, the PSD measured at zero current was subtracted from the data obtained at a given current flow to provide pure PSD of sample fluctuations.

Metastable resistivity states were induced by applying voltage pulses to the current contacts at liquid nitrogen temperatures. After a single pulse of 10 V and 5 sec duration the resistivity increased to what will be referred to as state 1. After performing series of relevant  measurements yet another state was enforced by applying a 30 V pulse to a sample in state 1 at low temperatures. This created yet another higher resistivity state that will be further referred to as the state 2.

\begin{figure}
\includegraphics*[width=7.5truecm]{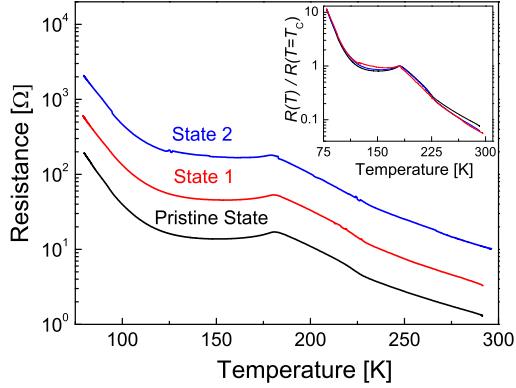}
\caption{Temperature dependence of the resistivity in the pristine state and in two subsequent metastable states, state 1 and state 2, created by application of voltage pulses at 77 K. Inset shows collapse of temperature dependencies of the resistivity normalized to the resistivity at $T=T_{\rm C}$ for all states into a single curve.}
\end{figure}

The temperature dependence of the sample resistance in three different states, pristine, state 1, and state 2, is shown in Figure 1. It is clear that the voltage pulse procedures shift the resistance up in entire investigated temperature range. One notes that characteristic temperatures in $R(T)$ curves, such as $T_S \approx 220$ K  associated with Jahn-Teller transition, temperature of the resistivity maximum at  metal-insulator transition temperature $T_{M-I}$ that in our samples coincides with Curie temperature $T_C \approx 180 $K, and temperature of the local resistivity minimum $T_{min}\approx 120$ K are not influenced by the electric pulse procedures. By plotting the resistance normalized by the resistance at $T_C$, one can collapse all $R(T)$ dependencies into a single $R(T)$ curve, as shown in the inset of Figure 1. The collapse means that even if the values of sample resistance increase after the pulse procedures, as shown in Fig. 1, the form of the $R(T)$ practically does not change. This behavior is markedly different from that of previously observed high resistivity metastable states, where remarkable resistivity differences between resistivities of different states were observed only at temperatures below $T_S$.\cite{PRB018,PRB020}  This is likely due to the difference in the pulse procedures employed. In earlier works a series of several short pulses has been used rather than a single long pulse employed here. Due to sample ageing we found it very difficult to enforce metestability by current sweeping or short pulses series, what forced us to use single long pulses instead. Long pulses may result in strong local overheating of the sample and in  micro cracking of the crystal. Long range strain fields induced by such cracks are known to be responsible for resistive properties of single crystalline manganites.\cite{V3} Appearance of micro cracks changes the underlying energy landscape for charge carriers and thus influences the nature of intrinsic tunneling at low temperatures. These changes can be revealed by low frequency noise measurements discussed below.

We have monitored changes of the resistance along with changes in the noise characteristics after creating different resistivity states. The noise spectral density was measured at seven temperatures chosen in such a way as to probe the noise properties in different magnetic states and with transport dominated by different mechanisms of dissipation. Namely, sample was checked in the paramagnetic insulating state, in the state close to metal-insulator transition associated with para to ferromagnetic transition, in the ferromagnetic percolating regime and in the low temperature regime with dominating intrinsic tunnel conductivity. We have found, as in our previous investigations, that at all measured temperatures the measured noise is of 1/f-type. The results of our noise measurements are summarized in Fig. 2, showing 3D plots of temperature and bias dependence of the noise PSD at 1 Hz and spectral exponent $\alpha$ obtained by fitting the spectra to $1/f^\alpha$ power law.

\begin{figure}
\includegraphics*[width=15truecm]{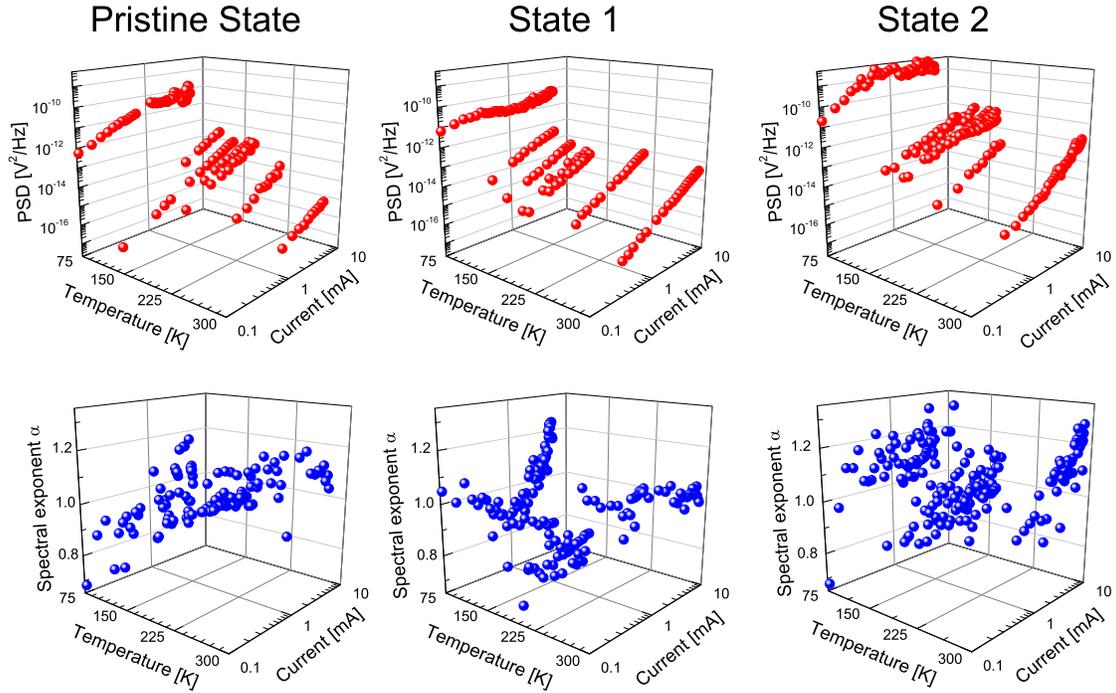}
\caption{Power spectral density at 1 Hz and spectral exponent of 1/f$^\alpha$ conductivity noise in different metastable states as a function of temperature and bias current.}
\end{figure}

We have previously verified that 1/f noise in our La$_{0.82}$Ca$_{0.18}$MnO$_3$ crystals obeys DH model.\cite{JAP} For such Gaussian noise all the information contained in the power spectral density of the noise should be fully equivalent to that contained in the dynamic resistivity. However, while the resistivity data nicely collapse to a single curve, the temperature dependencies of PSD of noise in different states remind to be markedly different. In addition, spectral exponents of noise in different states evolve with temperature in a pronouncedly different way. Apparently, the noise data are much more sensitive to small changes in the energy landscape than the resistance evolution is, and reveal these changes more pronouncedly.

Similarly to previous experiments, the noise observed in all investigated resistivity states of La$_{0.82}$Ca$_{0.18}$MnO$_3$  single crystals  has an equilibrium character at high temperatures and low bias, and its PSD scales as the square of the bias. Nevertheless, at temperatures below the resistivity upturn in all resistivity states the low bias "equilibrium 1/f noise"  turns into a "nonequilibrium" one above some threshold bias. The value of the threshold bias is different for each state, although the dependence seems not to scale monotonically with the resistivity of the sample. This can be clearly seen in Fig. 3 which shows the normalized PSD ($S_v/V^2$) as a function of bias voltage for different states. The normalized $S_v$ of an "equilibrium noise" should be constant and bias at which $S_v$ starts to decrease marks the threshold level above which the "nonequilibrium noise" is observed. The behavior of the measured normalized spectral density in Fig. 3 suggests that the noise turns back to an "equilibrium" one at high bias. This effect is most clearly visible in the  noise characteristic of state 1 where $S_v=S_V/V^2$ returns to be again bias independent for voltages exceeding 1 V.

\begin{figure}
\includegraphics*[width=7.5truecm]{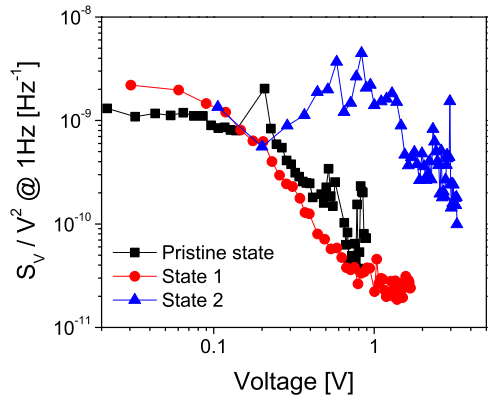}
\caption{Normalized noise power spectral density $S_v/V^2$ at 1 Hz as a function of bias voltage in different metastable states.}
\end{figure}

In our previous work we have ascribed the transition to "nonequilibrium 1/f noise" with changes in the intrinsic tunneling mechanism and underlying energy landscape. These changes were revealed by means analyzing bias dependence of the experimental second derivative of the I-V curve,  $dI^2/dV^2$, as in classical inelastic tunnel spectroscopy. Figure 4 shows experimental $dI^2/dV^2$ vs. $V$ characteristics of our sample in different resistivity states.

\section{Discussion and conclusions}
Let us analyze the conductance of La$_{0.82}$Ca$_{0.18}$MnO$_3$  single crystal at low temperatures as if being constituted by a large self-organized distributed tunnel junction. We justify such approach by noting that in GM theory of indirect tunneling the current due to hopping via localized states is  averaged across a large-area self organized junction. It is obvious that a set of parallel small-area contacts is equivalent to a single contact with a large area. The total conductance of a system of parallel contacts, obtained by summing over conductivities of individual junctions, is  in fact equivalent to the GM averaging procedure. From the first sight, the situation in a network of small contacts is series is different. In order to use our approximation we have to assume that the resistibility of individual junctions connected in series are not much different. Because of these restrictions we will not attempt to extract the barrier parameters from the experimental data but will limit ourselves to evaluation of the dominating tunneling mechanism in the junctions ensembles from the shape of the experimentally measured I-V curve.
\begin{figure}
\includegraphics*[width=7.5truecm]{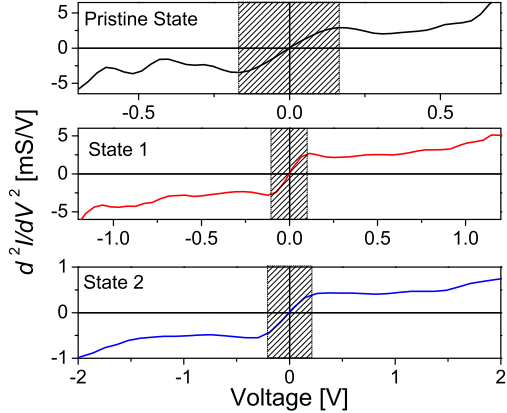}
\caption{Voltage derivative of the sample conductance measured at 77 K in different metastable states as a function of bias voltage. Hatched boxes indicate bias range at which the conductance is roughly proportional to $V^2$.}
\end{figure}

From $dI^2/dV^2$ characteristics of the sample in different resistivity states shown in Fig. 4 we see that the conductance  at very low voltages, marked by hatched boxes in the figure, is  proportional to $V^2$. For biases above the hatched boxes, and  below about 1 V, the conductance is roughly proportional to $|V|$. Note that we do not observe a linear contribution to $dI/dV$ of the form $const*V$ as in conventional metal-insulator-metal tunnel junctions, but a non analytical dependence $dI/dV = const*|V|$ at small voltages. This feature is especially clearly seen in the state 2 characteristics. Similar nonanalytic behavior of the tunnel conductivity background was already observed in many tunnel junctions involving perovskite manganites and high-$T_C$ superconductors. There was a wide discussion in the literature concerning the origins of this feature and it is now generally recognized that linear contribution to conductivity stems from inelastic tunneling from a broad continuity of states that are nearly equally distributed over the concerned energy scale. \cite{kirtley,LTP} Opening of each state for tunneling, at the bias corresponding to the given state energy, results in a step in differential conductivity $dI/dV$ at this specific bias. If there are many states which are uniformly distributed in energy, superposition of many corresponding steps gives a linear contribution to the differential conductance. If the structure is symmetrical and the excitations are within the barrier, the process is the same for both voltage polarities.

Existence of widely distributed localized states, whose formation is likely due to the structural disorder, within intrinsic barriers, enables strongly temperature dependent inelastic indirect tunneling across intrinsic junctions. The $|V|$ contribution should be  seen as a hallmark of the presence of such states and the domination of inelastic tunneling mechanism. The localized states in intrinsic barriers have profound effects on the tunneling characteristics. Inelastic temperature dependent tunneling conductivity in doped manganites appears as addition to spin-polarized elastic tunneling conductance.  At low temperatures and small bias these localized states may become resonant centers and enhance the tunneling probability. Increasing temperature or bias voltage enhances inelastic process due to, e.g., electron-phonon interaction. The dominating conduction paths in this regime are isolated chains containing localized states that are nearly rectilinearly and equidistantly positioned across the tunnel barrier. Inelastic hopping through these chains gives rise to the voltage and temperature dependence of tunneling conductance which is well described by the GM model. Characteristic of this process is that, as the voltage increases, the dominant contribution to the conduction comes from channels with larger and larger numbers of localized states. Convincing experimental evidence exists for the lowest order inelastic effects, and for crossover to higher order hopping channels with larger number of hops, with increasing bias.  As the barrier thickness, bias or the temperature are increased, the conductance crossovers into an asymptotic form which resembles the Mott Variable Range Hopping (VRH) behavior in the bulk limit.\cite{beasley} Clearly, for very low voltages, within hatched regions, the conductance is proportional to the square of the voltage and can be ascribed to direct or resonant tunneling processes. In the intermediate bias range where the GM inelastic tunneling dominates the conductance is a linear function of bias voltage. For voltages higher than the upper edge of the  uniformly distributed states band energy the conductance deviates from linear dependence on bias and the system crossovers from directed indirect tunneling regime to variable hopping regime.

By confronting the bias dependence of normalized PSD of noise at 77 K illustrated in Fig. 3 with the bias dependence of $dI^2/dV^2$ characteristics one notices that the voltage range of the linear conductance dependence on bias voltage coincides with the bias range at which "nonequilibrium" behavior of the noise is observed. Below this range the noise is clearly an "equilibrium" one and its PSD scales as the square of the bias. It seems that our data indicates also that the noise at high bias, above the upper edge of the energy of uniformly distributed localized states, returns back to "equilibrium" one. The conclusions regarding high bias behavior have to be taken however with the grain of salt. Experimentally, at high bias one faces strong dissipation in the sample that may lead to heating effects disturbing proper observations. From the theoretical point of view,  high-bias regime violates many of the assumptions of the models quoted above and therefore, a new approach is needed to account properly for the crossover from directed conduction along quasi-one-dimensional chains of localized states to the three-dimensional diffusive percolation of VRH. \cite{beasley,blanter}

\begin{figure}
\includegraphics*[width=7.5truecm]{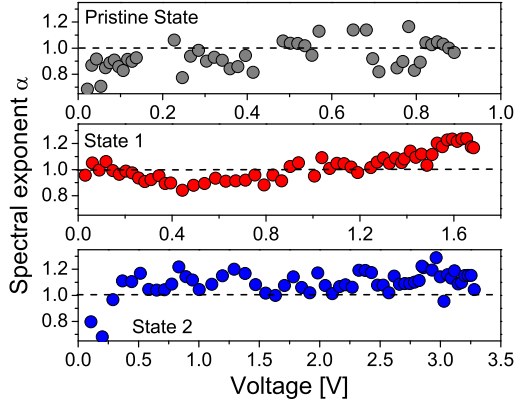}
\caption{Spectral exponent of the noise as a function of bias voltage in different metastable states at 77 K.}
\end{figure}

Our results suggest that "nonequilibrium" conductance noise seems to be directly related to the inelastic tunneling regime. Different properties of the noise in different metastable states result from changes in the structure of the indirect tunneling paths induced by the electric pulse procedures pulse. Therefore, changes in the noise should be directly linked to the form and shape of the energy landscape along the tunneling path. For fluctuations obeying the DH model of 1/f noise, the information of the energy landscape is contained in the behavior of the spectral exponent of the noise $\alpha$. We have shown in our recent paper \onlinecite{JAP} that noise in the investigated La$_{0.82}$Ca$_{0.18}$MnO$_{3}$ single crystal closely follows the DH behavior. Let us look therefore on the bias dependence of the spectral exponent in different metastable states at 77 K illustrated in Fig. 5. The most clear correlation between changes of the noise character and evolution of the exponent are seen in the characteristics of state 1. At low bias, in the equilibrium regime, $\alpha>1$ indicating existence of excess of  high energy states along the conduction path. At the threshold voltage, marking a crossover into "nonequilibrium" noise regime associated with change in the tunneling mechanism from a direct or resonant, into indirect inelastic one, the spectral exponent becomes smaller than 1 and stays smaller than 1 for the entire bias range in which $d^2I/dV^2\approx const$. One may claim that in this state, the "nonequilibrium" noise is observed when there is excess of low energy states along the conduction path. Noise returns to "equilibrium" behavior around 1 V and for bias $V>1$ V $\alpha$ becomes again higher than 1. However, the behavior of the spectral exponent in other states is not so evident and no general claim can be drawn from the data. Nevertheless, it seems that changes in noise character can be each time related to $\alpha$ crossing the $\alpha=1$ line.

Fluctuators responsible for 1/f noise in the tunneling regime can be most naturally associated with charge traps typically associated with structural defects inside intrinsic tunneling barriers. Such defects can be easily modified by electric field applied during the pulse procedures. Trapping and releasing of charge carriers from traps located inside tunnel barriers modulates the height of the barriers and leads to conductance fluctuations. Each trap constitutes an elementary two-level fluctuator for the 1/f noise. It is namely distribution of traps energies that is changed by pulse imposed changes in the resistivity states. Energy structure of such traps is modified by the applied bias what may lead to changes in the character of the noise with changing bias.

In conclusion, we have investigated  1/f noise in bias pulses induced metastable resistivity states in La$_{0.82}$Ca$_{0.18}$MnO$_{3}$ single crystal. Changes in the resistivity have been tentatively associated with microcracks developing in the crystal under pulse procedures. We have found that observed conductance noise turns over into a "nonequilibrium" one in the bias range at which the conductance is roughly proportional to $|V|$. This bias range has been associated with inelastic tunneling from a broad continuity of equally distributed states over the concerned energy scale. When conduction regime crosses over from inelastic tunneling to variable range hopping the noise seems to return to equilibrium. Data concerning bias dependence of the spectral exponent and bias enforced changes of the underlying energy landscape do not provide a comprehensive picture for all investigated states and require moire detailed further investigations.

\acknowledgments This research was supported by the Israeli Science Foundation administered by the Israel Academy of Sciences and Humanities (grant 754/09).

\end{document}